\providecommand{\doctitle}{Uncertainty Quantification of Super-Resolution Flow Mapping in Liquid Metals using Ultrasound Localization Microscopy}
\providecommand{\docauthor}{Weik et al.}

\documentclass[a4paper,11pt]{scrartcl}
\usepackage[utf8]{inputenc}
\usepackage[T1]{fontenc}

\usepackage{algorithmic}
\usepackage{textcomp}

\usepackage[english]{babel}
\usepackage[babel]{microtype}

\usepackage{todonotes}
\usepackage{url}
\usepackage{import}
\usepackage{nicefrac}
\usepackage{csquotes}

\usepackage{array}
\usepackage{booktabs}   
\usepackage{multirow}   
\usepackage{makecell}
\usepackage{xltabular}
\newcolumntype{L}[1]{>{\raggedright\let\newline\\\arraybackslash\hspace{0pt}}m{#1}}
\newcolumntype{C}[1]{>{\centering\let\newline\\\arraybackslash\hspace{0pt}}m{#1}}
\newcolumntype{R}[1]{>{\raggedleft\let\newline\\\arraybackslash\hspace{0pt}}m{#1}}

\usepackage{libertine}
\usepackage[libertine]{newtxmath}

\usepackage{siunitx}
\usepackage{mathtools}
\newcommand{\vect}[1]{\boldsymbol{#1}}

\usepackage{tikz}
\usetikzlibrary{calc}
\usepackage{pgf}
\usepackage{pgfplots}
\usepgfplotslibrary{groupplots}
\usepgfplotslibrary{external}
{main}

\usepackage{graphicx}
\usepackage{subcaption}
\usepackage{wrapfig,colortbl}

\usepackage{color}
\definecolor{ittengreen}{RGB}{0,142,91}
\definecolor{ittenblue}{RGB}{42,113,176}
\definecolor{ittenpurple}{RGB}{109,57,139}
\definecolor{ittenred}{RGB}{227,35,34}
\definecolor{ittenorange}{RGB}{241,142,28}
\definecolor{ittenyellow}{RGB}{224,229,0}

\usepackage[
  backend=biber,
  safeinputenc,
  hyperref=true,
  style=ieee,
  sorting=none,
  maxbibnames=20,
  maxcitenames=1,
  mincitenames=1,
  url=false,
  doi=false,
  isbn=false,
  eprint=true,
  backref=false,
  giveninits=true,
  date=year,
]{biblatex}
\usepackage{hyperref}
\hypersetup{
    hidelinks,
    colorlinks=false,
    bookmarksnumbered=true,
    pdftitle=\doctitle,
    pdfauthor=\docauthor,
}

\usepackage{authblk}

\begin{document}

\title{\doctitle}
\date{}

\author[1]{David Weik}
\author[1]{Zehua Dou}
\author[2]{Dirk Räbiger}
\author[2]{Tobias Vogt}
\author[2]{Sven Eckert}
\author[1]{Jürgen Czarske}
\author[1]{Lars Büttner}

\affil[1]{Laboratory of Measurement and Sensor System Techniques, TU Dresden, 01069 Dresden, Germany}
\affil[2]{Department of Magnetohydrodynamics, Helmholtz-Zentrum Dresden-Rossendorf, 01314 Dresden, Germany}

\maketitle

\begin{abstract}
Convection of liquid metals drives large natural processes and is important in technical processes.
Model experiments are conducted for research purposes where simulations are expensive and the clarification of open questions requires novel flow mapping methods with an increased spatial resolution.
In this work, the method of Ultrasound Localization Microscopy (ULM) is investigated for this purpose.
Known from microvasculature imaging, this method provides an increased spatial resolution beyond the diffraction limit.
Its applicability in liquid metal flows is promising, however the realization and reliability is challenging, as artificial scattering particles or microbubbles cannot be utilized.
To solve this issue an approach using nonlinear adaptive beamforming is proposed.
This allowed the reliable tracking of particles of which super-resolved flow maps can be deduced.
Furthermore, the application in fluid physics requires quantified results.
Therefore, an uncertainty quantification model based on the spatial resolution, velocity gradient and measurement parameters is proposed, which allows to estimate the flow maps validity under experimental conditions.
The proposed method is demonstrated in magnetohydrodynamic convection experiments.
In some occasions, ULM was able to measure velocity vectors within the boundary layer of the flow, which will help for future in-depth flow studies.
Furthermore, the proposed uncertainty model of ULM is of generic use in other applications.
\end{abstract}

\section{Introduction}
\label{sec:introduction}
\subsection{Motivation}
Fluids with very low Prandtl numbers are characterized by very high thermal conductivity and low viscosity at the same time.
They are relevant for geophysical and astrophysical convection flows such as liquid metal convection in the liquid core of the Earth or the convection zone of the Sun.
This low Prandtl number convection is numerically challenging, as the magnetohydrodynamic nature of liquid metals introduces a further force and very small simulation grids are required.
Typically, they can only be investigated in the laboratory using liquid metal experiments.
A useful paradigm to experimentally study these turbulent flows is the configuration of a Rayleigh-Bénard convection (RBC).
Using a low melting point alloy such as Gallium-Indium-Tin (GaInSn), these experiments can be scaled to a size and parameter range that can be realized in the laboratory.
Flow mapping close to the boundary layer in these experiment's dimensions can a breakthrough for the understanding of the turbulent flow phenomena.

\subsection{State of the Art}
Laser-based flow metrology can in principle provide the required resolution to study small scale dynamics near or within the viscous boundary layer \cite{Adrian2011,Bilsing2022} and recently contributed to the research of convection flows \cite{Kaeufer2023}.
However, optical waves cannot penetrate opaque liquid, such as liquid metals.
X-ray tomography was shown to enable the spatial and temporal resolution \cite{Barthel2015,Moreno2023}.
But still, the penetration depth is a limitation, as the absorption coefficient of metallic materials is too large.
Neutron radiography showed large penetration depths \cite{Takenaka1996,Lappan2021,Skrypnik2024} but it can only track comparably large particles due to its low sensitivity and can therefore not provide the required spatial resolution.
\par
Another way for flow metering is to make use of the liquid metals magnetohydrodyamic nature: If the convection flow is operated within a constant magnetic field, any movement will induce additional magnetic field strengths.
Based on this principle, \textcite{Stefani2000} initially proposed the method of conductive inductive flow tomography (CIFT) for detecting the flow using fluxgate sensors.
This non-contact method was demonstrated for observations in continuous steel casting \cite{Wondrak2010} and recently enabled a macroscopic and three dimensional flow observation within a large convection cell \cite{Wondrak2023}.
Despite CIFT's promising results, its spatial resolution (\qty{20}{\mm}) is too low for measurements close or within the boundary layer.
\par
In recent years, the modality of ultrasound imaging was shown to provide promising capabilities, which were firstly demonstrated by \textcite{Takeda2002}.
Intrinsic oxides and impurities of the material act as scattering particles for ultrasound \cite{Wang2021} of which the velocity vectors can be distinguished.
Previous works enabled multi-dimensional multi-component vector Doppler imaging in liquid metal flows \cite{Thieme2019b,Nauber2020}.
Sequential scanning allows the observation of large areas, such as a $\qty{200}{\mm}\times\qty{200}{\mm}$ cross-section shown by \textcite{Thieme2019}.
\par
To further increase the spatial and temporal resolution and likewise use only a single access, phased array plane wave scanning is of particular interest \cite{Udesen2008, Montaldo2009}.
In a previous work \cite{Weik2021}, this method was combined with ultrasound image velocimetry (UIV) \cite{Poelma2012, Leow2015, Poelma2017} and introduced for liquid metal flow mapping.
UIV was shown to be capable of resolving the temporal dynamics of the convection flow and enable modal analysis \cite{Weik2021b}.
But with spatial resolution of \qty{4}{\mm}, it was limited to the larger main structures.
The re-circulation areas, near-wall flow profiles and detached areas remain hidden from UIV imaging.
\par
For the deeper understanding in these structures, a further increase of spatial resolution is of particular interest.
Here, the method of super-resolution ultrasound localization microscopy (ULM) can introduce a meaningful contribution \cite{Siepmann2011, Desailly2013, Ackermann2015, Diamantis2018}.
Given the certainty of an echo created from a single scatterer, its center position can be estimated within a resolution that corresponds to a fraction of the wavelength.
In blood flow imaging, such single scatterer information is provided by using nonlinear contrast agents.
Their echoes can be separated from other reverberating structures by their harmonic response \cite{Bouakaz2002}.
This way, flow profiles can be measured on a small scale, which enables the investigation of the microvasculature \cite{Errico2015, Christensen2020, Demene2021, Renaudin2022} or small channels \cite{Kupsch2021}.
\par
For liquid metal imaging however, single isolatable scatterers are not easy to capture.
Contrast agents that overcome the high surface tension and that have a similar density to the liquid metal are not readily available.
Furthermore, as reported in earlier studies \cite{Wang2021, Thieme2019}, a homogeneous distribution of the intrinsic scattering particles in the experiments is difficult to achieve due to sedimentation processes.
The experiments are typically conducted as follows:
\begin{itemize}
 \item Initial mixing of the melt,
 \item Relaxation time of \qtyrange{30}{60}{\minute} to ensure unbiased flow conditions,
 \item Test of B-mode quality,
 \item Start of measurement.
\end{itemize}
In general, applying the processing method of ULM without the pre-condition of isolatable scatterers will introduce measurement uncertainties.
A solution to apply ULM in liquid metal flow studies was firstly demonstrated in a previous work by applying nonlinear adaptive beamforming \cite{Weik2022}.
With this method, a spatial resolution of \qty{188}{\micro\m} was proposed.
Despite this promising resolution, as ULM in liquid metal flows is a completely new modality, the uncertainty of the velocity vectors were yet not investigated, nor was this method shown to be reproducible.
Also, a validation against a traceable reference was yet not shown.
This is crucial to provide trustworthy results for the investigation of flow regimes in future convection studies that go beyond the known parameter range.

\subsection{Aim and Outline}
As a novelty of this work, a generic model to quantify the uncertainty of flow maps obtained with ULM is investigated.
This is interesting to generate quantified flow maps with ULM but of particular interest in magnetohydrodynamic (MHD) research, as new turbulent flow phenomena in GaInSn can be investigated that cannot be referenced otherwise.
\par
In this article, we will firstly explain how a ULM system was adapted for the targeted convection experiments in \autoref{sec:methods}.
In \autoref{sec:uncertainty} the uncertainty estimation model is introduced and a validation will be conducted against a reference.
Finally in \autoref{sec:application}, demonstrations of ULM in MHD studies are shown.

\section{Adaptation of Ultrasound Localization Microscopy}\label{sec:methods}
\subsection{Nonlinear Adaptive Beamforming}
\begin{figure}[!bt]
    \centerline{%
        \includegraphics[width=\linewidth]{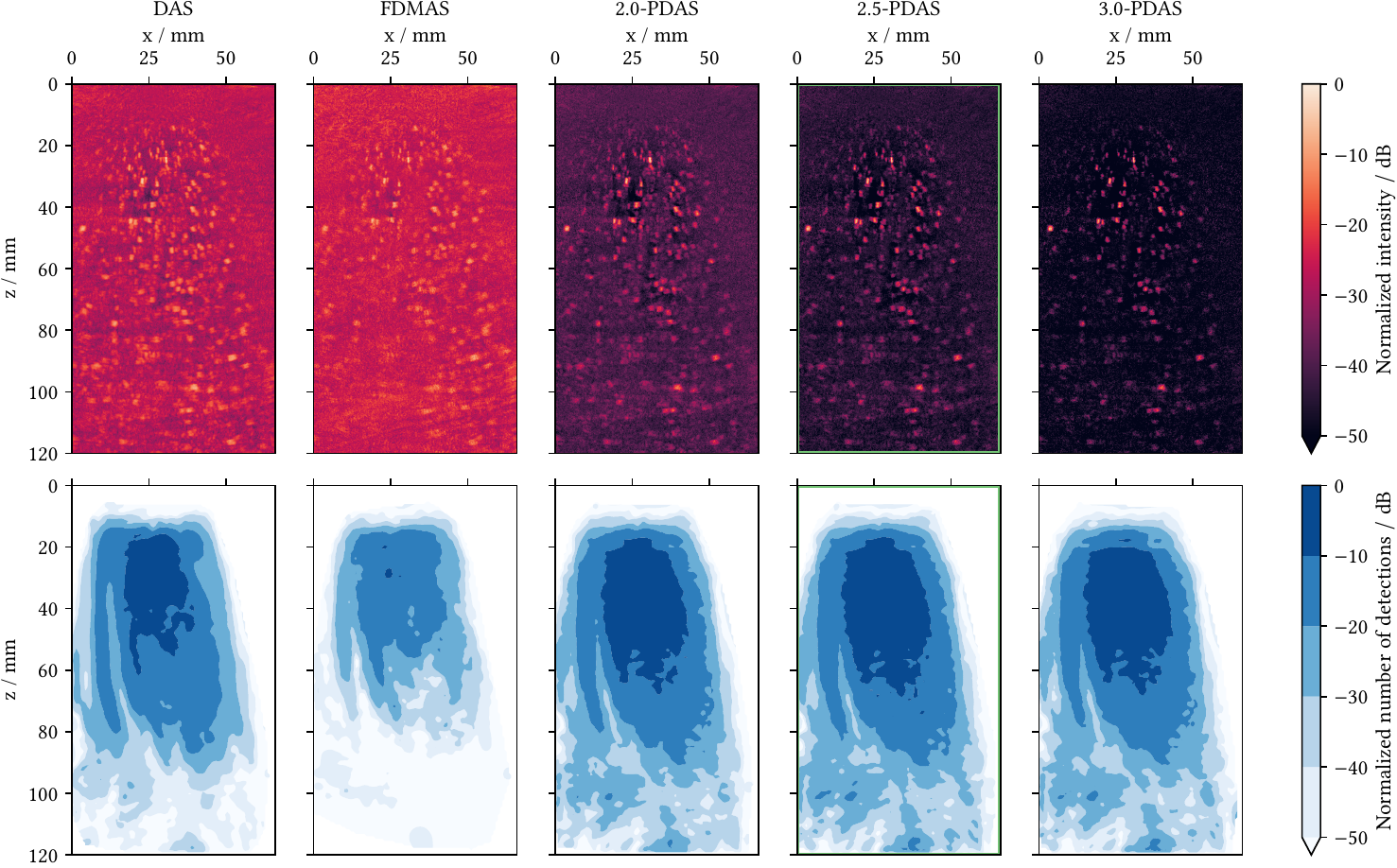}
    }
    \caption{
    Application of various beamformers for the scatterer imaging and tracking in the melt flow.
    In the top, the B-mode images is shown after SVD-filtering.
    On the bottom line, a map containing the number of individual particle detections  is shown for each beamformer after tracking and filtering.
    Left, the standard DAS image is shown were single scatterers cannot be isolated due to the high density.
    FDMAS sharpens and suppresses some tracers, however, no sufficient fine tuning can be applied and even less tracers are detected as with DAS.
    In comparison, the PDAS showed the best results.
    Depending on the $P$-value, minor echoes are supressed resulting into images with increased contrast.
    For this example, the value of $P=\num{2.5}$ showed the highest number of detections.
    }
    \label{fig:ulm_bmode}
\end{figure}

In this work, the scalable nonlinear beamforming method of $P^{\mathrm{th}}$-root compressed delay-and-sum (PDAS) was applied to image the scattering particles \cite{Polichetti2018}.
In \autoref{fig:ulm_bmode} example B-mode images from a liquid metal convection experiment are shown, where different scalings of PDAS are compared to conventional delay-and-sum beamforming (DAS) and another nonlinear beamforming method of filtered delay-multiply and sum (FDMAS) \cite{Matrone2015}.
DAS beamforming produces too high a concentration of scattering particles and too high a speckle intensity as to ensure a minimal tracking uncertainty.
This is a result of the varying scattering particle concentration that is not easily adjustable during the experiment.
The point spread function, especially for penetration depths of $>\qty{50}{\mm}$, is not sufficiently small to isolate and track the particles.
\par
FDMAS was a first nonlinear beamforming approach that is based on coherence weighting \cite{Matrone2015}.
Due to the multiplication step in the beamforming algorithm, structures are emphasized that are coherent in other channels signals.
This operation therefore comprises a nonlinear transfer function and was shown to suppress speckles and enhance the lateral point-spread-function \cite{Matrone2019}.
As shown in \autoref{fig:ulm_bmode}, FDMAS was not beneficial for the particle images investigated.
The contrast is reduced and the resolution is not improved compared to DAS.
For penetration dephts beyond \qty{60}{\mm} the resolution is even less.
In \autoref{fig:ulm_bmode} a map containing the number of valid detections of the tracking algorithm is given for each beamformed set of B-modes.
Here, FDMAS is also not shown to be beneficial.
A possible explanation for this result can be that the materials temperature is not static and highly locally variable throughout a convection experiment.
This therefore also accounts to the speed-of-sound and to the delays.
The coherence weighting of FDMAS can be sensitive for these fluctuations and a supression of particle echoes that have a varying delay.
\par
Compression beamforming is more prone to delay fluctuations as used in wireless communications and introduced to ultrasound imaging by \textcite{Polichetti2018} as $P^{\mathrm{th}}$-root compressed delay-and-sum (PDAS).
The echo signals obtained from each emitted plane wave tilt are compressed by their $P^{\mathrm{th}}$-root, DAS beamformed and backward raised to the power of $P$.
Afterwards, each plane wave tilt is added.
Compared to FDMAS it employs root compressed autocorrelation.
It produces comparable results, has a better performance with speed-of-sound fluctuations, a higher computational efficiency and the grade of the nonlinear scaling is arbitrarily adjustable by the value $P$.
Furthermore, PDAS allows for the adaptive scaling of the beamforming algorithm to the actual particle concentration.
In \autoref{fig:ulm_bmode} the benefit of this approach compared to DAS and FDMAS is shown.
It can be seen, that the particle profile sharpened as the value of $P$ increased.
Additionally, less intense scatterers are further suppressed, which is helpful for the tracking in an increased particle density.
On the other hand, a loss of information in the corner regions and higher penetration depths were seen for values of $P>\num{2.5}$.
\par
\begin{figure}[ht]
    \centering
    \includegraphics{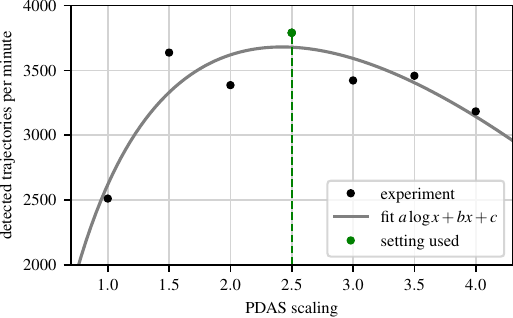}
    \caption{
        Detected trajectories per minute from the example convection dataset depending on the $P$-value of the PDAS beamformer.
    }
    \label{fig:pdas_traj}
\end{figure}

For each experiment, the tracking performance was judged to identified optimal $P$-values.
The PDAS was adjusted individually to compromise the number of individual trajectories, the average trajectory length and the found trajectories in the corner regions and higher penetration depths.
The number of valid detections depends on the $P$-value in each individual experiment, as demonstrated in \autoref{fig:ulm_bmode}.
As a generic metric, the average number of trajectories per minute was calculated for each $P$ value, see \autoref{fig:pdas_traj}.
For the example experiment, $P=2.5$ showed the best metric.
Depending on the particle density, typical values were in the range of $P=2.5\pm0.5$.
Additionally as a supplement to this paper, a video of this B-mode sequence is provided with switching values of $P$.

\subsection{Particle Filtering, Localization and Tracking}
In the experiments, the images will be captured through a \qty{3}{\mm} acrylic glass window, which will introduce severe clutter for up to \qty{20}{\mm}.
The clutter was filtered using a singular value decomposition (SVD) filter \cite{Demene2015,Song2017,Hasegawa2019}.
From a stack of 400 frames, the first singular vector and the ones after 250 were neglected.
Using less vectors was not applicable as it led to a smearing of particles and a spread of distortions that came from other ultrasound probes over multiple frames.
\par
For the localization and tracking of the scattering particles, the software package \textit{PtvPy}
was used.
\textit{PtvPy} provides an integrated software tool that is based on the \textit{Python} package \textit{trackpy} \cite{trackpy}, in which the localization and linking of particle positions is conducted according to \textcite{Crocker1996}.
First, the background image is calculated and subtracted from each frame.
Second, the centroid method is applied for the localization of the center of each particle from the B-Mode images \cite{Couture2018}.
Third, the identified particles are linked to trajectories through the estimation of the most probable location between consecutive frames using the nearest neighbor method.
Trajectories that were followed for at least \qty{250}{\ms} are neglected.
The velocity vector $\vect{v}$ at the index $n$ of a trajectory is calculated with the difference from one localization with the position vector $\vect{r}$ to the next, divided by the step time:
\begin{equation}\label{eq:ulm_velocity}
    \vect{v}_{n} = \frac{\vect{r}_{n+1} - \vect{r}_{n}}{\Delta t_{n}}.
\end{equation}
The step time $\Delta t_{n}$ can be the time for one to a multiple of frames, whether the algorithm detected a re-appearance.
Here, a memory of three frames was used until a localized particle can vanish and reappear nearby.
This was done to compensate for fluctuations due to noise.
The detection step size multiplied by the frame rate gives the step time
\begin{equation}
    \Delta t_{n} = \left(M_{n+1} - M_{n} \right) \cdot f_{\mathrm{frame}},
\end{equation}
where $M$ denotes the frame number of the $n$-th detection in the respective trajectory.
Finally, the mean velocity is calculated from the trajectories that have irregular coordinates by an interpolation to a regularly spaced grid.
For this purpose, the function \textit{scatter\_to\_regular} of \textit{PtvPy} was used%
.
The function was extended to provide either an interpolation of the mean, median, count or standard deviation value.
The interpolation of the mean velocity vector results in a 2D-2C flow velocity map.
An overview of the parameters used is shown in \autoref{tab:ulm_parameters}.
\begin{table}[!h]
    \centering
    \caption{
    Specifications used for ULM imaging.
    }
    \label{tab:ulm_parameters}
    \begin{tabularx}{\linewidth}{XX}
        \toprule
        \multicolumn{2}{c}{\textit{Measurement}}\\
        probe & 1.5-D phased array with 256 elements, \qty{4}{\MHz}, \qty{0.3}{\mm} pitch, PMMA matched\\
        excitation pulse & 2 periods at \qty{4}{\MHz} with 3 levels\\
        electrical amplitude & $\pm\qty{50}{\V}$\\
        pulse repetition frequency & \qty{1350}{\Hz}\\
        number of exc. angles & \num{7}\\
        angular range & $\pm\qty{30}{\degree}$\\
        sampling rate & \qty{20}{\MHz}\\
        downsampled frame rate & \qty{40}{\Hz}\\
        constant amplified gain & \qty{61}{\dB}\\
        time dependent gain & \qty{100}{\dB\per\ms}\\
        \midrule
        \multicolumn{2}{c}{\textit{Beamforming}}\\
        beamformer & PDAS with sub-sample interpolation\\
        apodization & PW angle dependent and single element directivity\\
        clutter filter & SVD\\
        \midrule
        \multicolumn{2}{c}{\textit{Velocity estimation}}\\
        algorithm & \textit{PtvPy}
        \\
        expected particle spread & \qty{1.5}{\mm}\\
        minimum trajectory length & \num{10} frames\\
        trajectory linking memory & \num{3} frames\\
        maximum travel distance between frames & \qty{1.02}{\mm}\\
        maximum velocity & \qty{40.7}{\mm\per\s}\\
        \bottomrule
    \end{tabularx}
\end{table}

\section{Uncertainty Quantification}\label{sec:uncertainty}
\subsection{Velocity Uncertainty}
The overall velocity uncertainty of ULM in this application comprises of three main components: localization uncertainty, velocity gradient as well as flow and temperature fluctuations.
The square sum of all components gives the estimate of the overall measurement uncertainty as
\begin{equation}\label{eq:ulm_uncertainty}
    \sigma_{\vect{v}} = \sqrt{{\sigma_{\vect{v}, \Delta\vect{r}}}^2 + {\sigma_{\vect{v}, \nabla\vect{v}}}^2 + {\sigma_{\vect{v}, s}}^2}.
\end{equation}
An overview of the components and their determination is given in \autoref{tab:ulm_uncertainty}.
\begin{table}[!t]
    \centering
    \caption{
    Uncertainty quantification: GUM-type determination, model and typical estimate value.
    \label{tab:ulm_uncertainty}
    }
    \begin{tabularx}{\linewidth}{cXXcr}
        \toprule
        Quantity & Description & Determination & Model & Rel. average\\
        \midrule
        $\Delta\vect{r}$ & Velocity uncertainty due to a deviation of the particle localization & Type B: Calibration and propagation from \autoref{eq:ulm_velocity} & $\sqrt{\frac{2}{N_{\vect{r}}}} f_{\mathrm{frame}} \sigma_{\vect{r}}$ & \qty{2.92}{\percent}\\
        $\nabla\vect{v}$ & Velocity uncertainty due to a velocity gradient & Type B: Calibration and estimation \cite{Westerweel2008,Weik2021} & $\frac{\Delta\vect{r}}{2\sqrt{6}} \nabla\vect{v}$ & \qty{0.61}{\percent}\\
        $\vect{v}$ & Velocity uncertainty due to fluctuations of the flow and temperature & Type A: Average standard deviation from velocity data & $\frac{s_{\vect{v}}}{\sqrt{M}}$ & \qty{2.98}{\percent}\\
        \midrule
        $\vect{v}$ & Overall measurement uncertainty & Square sum of components & $\sqrt{\sum_{n}{\sigma_{\vect{v},n}}^2}$ & \textbf{\qty{4.22}{\percent}}\\
        \bottomrule
    \end{tabularx}
\end{table}
Each component is further explained in the following.
\par
ULM incorporates a localization uncertainty due to noise and non-uniform shapes of particles.
This defines how well the particles' center of gravity can be determined.
Its impact to the measured velocity at a specific position vector $\vect{r}$ can be derived by an error propagation through \autoref{eq:ulm_velocity} as
\begin{equation}
   \sigma_{\vect{v}, \Delta\vect{r}} = \frac{\partial \vect{v}}{\partial \vect{r}} \sigma_{\vect{r}}.
\end{equation}
The vector $\vect{v}$ comprises two individual localizations.
Therefore, the detected particle in both frames with their respective $\sigma_{\vect{r}}$ is considered individually and square summed.
Furthermore as a worst case scenario, it is considered that the step time $\Delta t$ is minimal with $1/f_{\mathrm{frame}}$.
Then, $\sigma_{\vect{v},\vect{r}}$ is estimated as
\begin{equation}
    \sigma_{\vect{v},\vect{r}} = \sqrt{\frac{2}{N_{\vect{r}}}} f_{\mathrm{frame}} \sigma_{\vect{r}},
\end{equation}
where $N_{\vect{r}}$ denotes the number of found trajectories at this position, $f_{\mathrm{frame}}$ the frame rate of the B-mode images and $\sigma_{\vect{r}}$ the localization uncertainty of an individual particle.
\par
Another influence on the local uncertainty in flow mapping is the velocity gradient.
Observing a velocity gradient within a limited spatial resolution results into a systematic deviation.
This is similar for ULM as for other velocity estimation methods such as particle image velocimetry (PIV) \cite{Westerweel2008} or its pendant ultrasound image velocimetry (UIV) \cite{Poelma2017,Weik2021}.
From these considerations, the uncertainty due to a velocity gradient $\sigma_{\vect{v}, \nabla\vect{v}}$ can be estimated by:
\begin{equation}\label{eq:unc_gradient}
    \sigma_{\vect{v}, \nabla\vect{v}} = \frac{\Delta\vect{r}}{2\sqrt{6}} \nabla\vect{v}.
\end{equation}
\par
At last, fluctuations of the flow introduce a random deviation of the derived velocities.
This also accounts to fluctuations of the temperature during the experiment, which alter the speed of sound and therefore shift the detected particle localization.
This random deviation is derived by estimating the standard deviation from the trajectories that contributed to each individual spot and normalized to the square root of the number of trajectories $M$:
\begin{equation}
    \sigma_{\vect{v}, s} = \frac{s_{\vect{v}}}{\sqrt{M}}.
\end{equation}

\subsection{Spatial Resolution and Localization Uncertainty}
Two components of \autoref{eq:ulm_uncertainty} depend on ULMs spatial resolution and localization uncertainty.
Therefore, these values need to be characterized within realistic experimental conditions.
For ULM, the spatial resolution is characterized by how well the center position of a single scatterer can be isolated.
In other words, spatial resolution and localization uncertainty are the same quantity.
In general, the localization uncertainty at a specific position $\vect{r}$ consists of a systematic uncertainty due to the finite size and uniform shape of the particle profiles and image distortions, as well as a random uncertainty due to noise:
\begin{equation}\label{eq:ulm_loc}
    \sigma_{\vect{r}} = \sqrt{ {\sigma_{\vect{r},\mathrm{syst.}}}^{2} + {\sigma_{\vect{r},\mathrm{rand.}}}^{2} }.
\end{equation}
In previous works, $\sigma_{\vect{r}}$ was characterized by means of a reference setup with a single scatterer \cite{Kupsch2021,Weik2022}.
The setup and the used probe were the same as with an initial study \cite{Weik2022}.
Therefore, the calibrated value value $\sigma_{\vect{r}}=\qty{188}{\micro\m}$ is also valid in this work for the localization uncertainty, i.e. the spatial resolution.
This equals \num{0.28} wavelengths in the medium.

\subsection{Validation}
\begin{figure}[!b]
    \centering
    \includegraphics{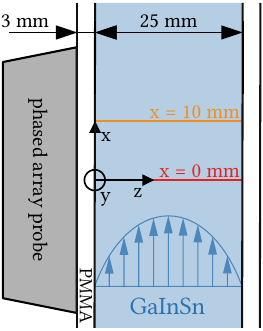}
    \caption{
    Cross-section of the reference flow setup.
    It provides a laminar flow that was referenced by a PIV measurement of a water flow and Reynolds-scaling.
    The lines indicated the profile positions shown in \autoref{fig:ulm_vel_ref}.
    }
    \label{fig:ulm_vel_ref_setup}
\end{figure}

So far, ULM in liquid metals was yet not validated against a reference.
X-Ray or Neutron imaging as a reference can be used in principle, but they take a high instrumental effort in the laboratory use.
We therefore choose a method that is based on Reynolds-scaling \cite{Weik2021}: A flow setup was built, that ensures laminar flow conditions within a $\qty{25}{\mm}\times\qty{40}{\mm}$ channel.
First, a water flow is introduced in the channel that can be observed with optical PIV.
Then, the water is exchanged with GaInSn.
The GaInSn is pumped with approx. three times less speed, which exactly reproduces the same Reynolds number as the water flow.
The PIV profile is then scaled according to the changed flow rate and can serve as a traceable reference for the ultrasound measurement in the channel.
A sketch of the setup is shown in \autoref{fig:ulm_vel_ref_setup}.
\par
\begin{figure}[!b]
    \centering
    \import{plots/ulm}{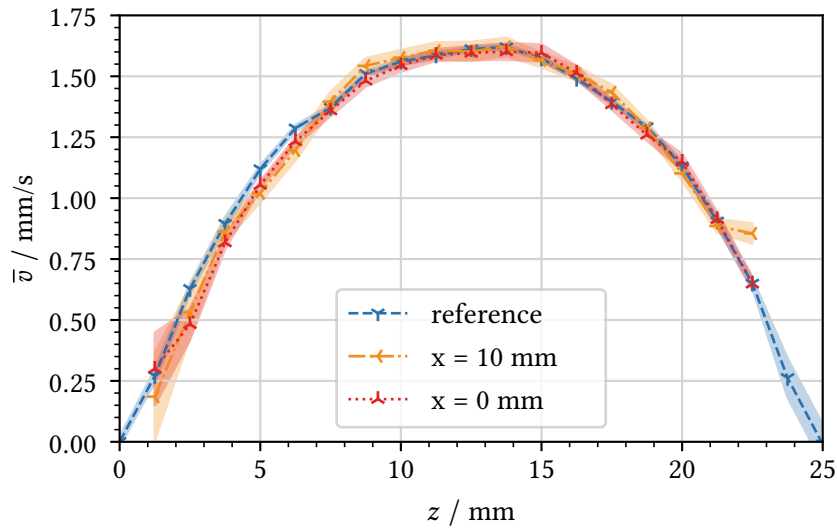}
    \caption{
    Validation of the ULM profiles.
    Two profiles of the ULM measured flow map are shown against the reference.
    The curves give the lateral velocity component along the axial direction, where the red curve is in the middle of the probe and the orange curve with \qty{10}{\mm} lateral displacement.
    The colored bars indicate the estimated uncertainty range.
    }
    \label{fig:ulm_vel_ref}
\end{figure}

From the beamformed dataset with \qty{1}{\minute} duration, \num{5819} unique trajectories of particles were identified.
These trajectories were averaged onto a regular grid of which two profiles were extracted, one in the middle of the transducer array ($x=\qty{0}{\mm}$) and one with a lateral displacement ($x=\qty{10}{\mm}$), respectively.
The profiles are shown in \autoref{fig:ulm_vel_ref} against the reference.
The uncertainty bars are estimated with the uncertainty budget of \autoref{eq:ulm_uncertainty}.
\par
Compared to earlier results that were acquired with ultrasound image velocimetry (UIV) in this channel \cite{Weik2021}, three major advancements of ULM come at hand:
\begin{itemize}
 \item
    The flow profiles at \qty{0}{\mm} and \qty{10}{\mm} are almost identical.
    The results with lateral offset incorporate much less systematic deviation.
    Lateral distortions are typical for phased array images.
    The ULM processing seems more prone against them compared to the interrogation area processing in UIV.
 \item
    Overall, the uncertainty-bars became smaller.
    For UIV \cite{Weik2021} this was only the case in the middle, were no velocity gradient was present.
    This is likely a result of the increased spatial resolution.
    Furthermore, an increased measurement time with more identified trajectories should decrease the uncertainty further.
 \item
    The systematic deviation with increased velocity gradient and closer to the walls is much less.
    This is also due to the increased spatial resolution.
    As a result, larger velocity gradients are observable with ULM.
\end{itemize}
Despite these benefits, a remaining issue are reflections from the front- and backwall that shadow the particle information.
No valid trajectory was tracked from \qty{1}{\mm} distance of the front wall or \qty{2}{\mm} distance to the backwall, even though SVD filtering was applied.
This shadowing effect can be reduced by using further attenuating wall materials.
\par
As an outcome of the uncertainty modeling it is shown that the contribution due to the velocity gradient is comparably small, as with an average estimate of \qty{0.61}{\percent} for typical velocity gradients.
Main influential is the number of found trajectories as well as their length.
This influences the main uncertainty contributions of particle localization and flow fluctuations, with each approx. \qty{3}{\percent} contribution.
Wall regions encounter less valid detections and must be considered with an increased uncertainty of approx. \qtyrange{10}{20}{\percent}, depending on the captured B-mode quality and the turbulence grade.

\section{Application in a Liquid Metal Convection Experiment}\label{sec:application}
\subsection{Rayleigh-B\'enard Convection Setup}
\begin{figure}[!b]
    \centering
    \includegraphics[width=\linewidth]{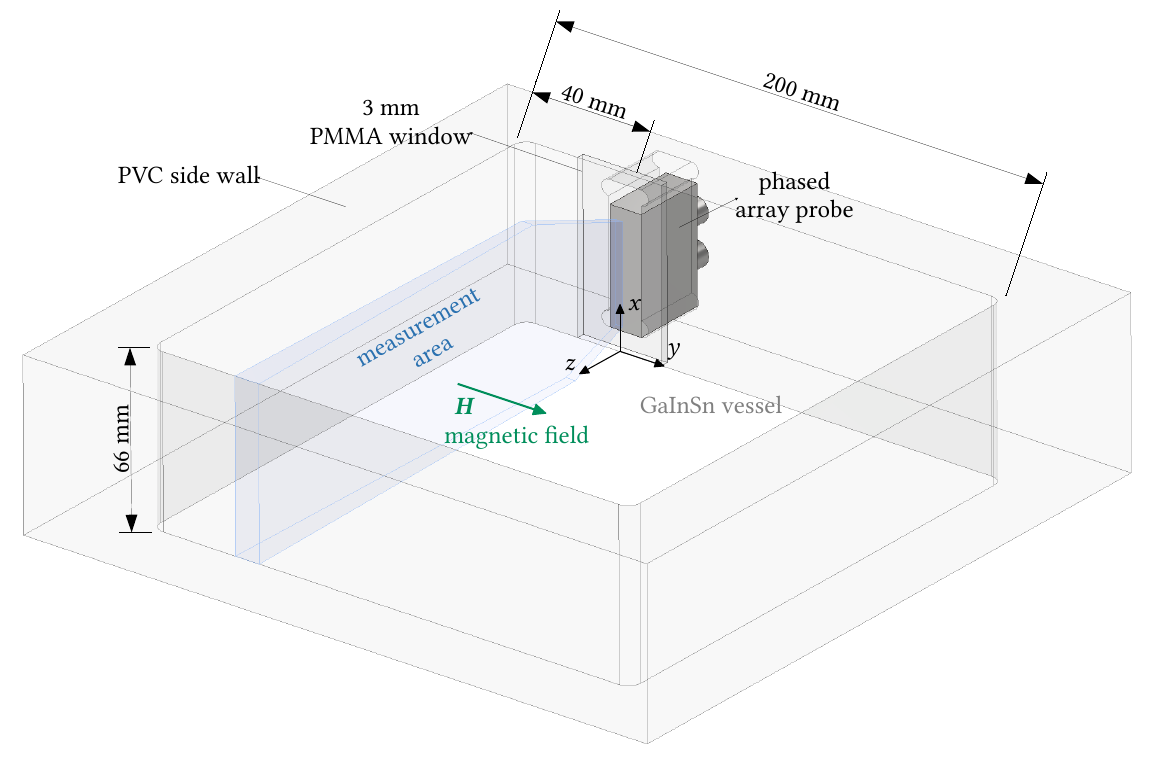}
    \caption{
    Designed container for RBC model experiments.
    Indicated are the dimensions, materials and the position of the phased array probe and the measurement area inside of the vessel.
    On top and on the bottom are the encasing heating and cooling plates, which are not shown here.
    }
    \label{fig:rbc_dirk}
\end{figure}

In this section, ULM is demonstrated to acquire flow vectors in laboratory liquid metal experiments.
The convection cell for this study was designed at the Magnetohydrodynamics Lab at the Helmholtz-Zentrum Dresden-Rossendorf \cite{Vogt2018}, see \autoref{fig:rbc_dirk}.
The container had a square cross-section with a side length of $L = \qty{200}{\mm}$ and a height of $H = \qty{67}{\mm}$.
This corresponded to an aspect ratio of $\varGamma = L /H = 3$.
The fluid layer was located between two temperature-controlled copper plates on the top and bottom of the cell, in which there was a network of channels that were connected to a temperature-controlled water circulation system.
The strength of the thermal forcing is described using the dimensionless Rayleigh number $\mathrm{Ra}$:
\begin{equation}
  \mathrm{Ra} =  \frac{\alpha g \Delta T H^3}{\kappa \vartheta},
\end{equation}
where the numerator is given by the fluid's thermal expansion $\alpha$ gravity $g$ as well as the temperature difference $\Delta T$ and the distance of the plates $H$.
The denominator is given by the fluid's thermal diffusivity $\kappa$ and kinematic viscosity $\vartheta$.
During the experiment, $\mathrm{Ra}$ was controlled by thermostatic baths and $\Delta T$ was observed by nine thermocouples in each plate.
The side walls were made of electrical non-conductive polyvinyl chloride.
The influence of heat loss and environmental temperature conditions was reduced by encasing the entire convection cell in \qty{30}{\mm} insulating foam.
The fluid vessel was filled with GaInSn as the model fluid.
At room temperature, it has a density of \qty{6.44}{\g\per\cubic\cm}, a viscosity $\vartheta$ that was approx. three times lower than water \qty{3.4e-7}{\square\m\per\s} and a thermal diffusivity $\kappa$ of \qty{1.05e-5}{\square\m\per\s}.
Accordingly, the Prandtl number is $\mathrm{Pr}=\vartheta/\kappa=\num{0.03}$.
The convection cell was placed between two electromagnetic coils to provide a horizontal magnetic field.
The spatial in-homogeneity of the magnetic field was less than \qty{5}{\percent} at maximum strength \cite{Tasaka2016}.
Different flow patterns appeared in the liquid metal convection depending on the ratio between the Rayleigh-number $\mathrm{Ra}$ and the Chandrasekhar number $Q$, which quantifies the effect of the magnetic field in the flow field.
The ultrasound array probe was placed upright with \qty{40}{\mm} offset from the side, as depicted in \autoref{fig:rbc_dirk}.
The measurement area of the probe was aligned perpendicularly to the magnetic field lines.
For a sufficient large magnetic field strength, the flow becomes 2D with respect to the magnetic field lines.

\subsection{Main Roll Deformation}\label{sec:application_main_roll}
\begin{figure}[!t]
    \centering
    \includegraphics{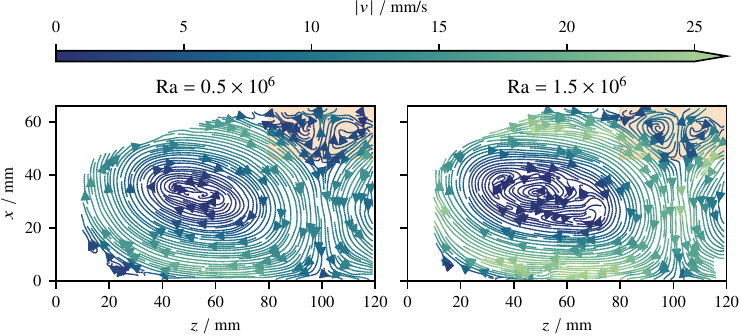}
    \caption{
    Measured flow field averaged over \qty{10}{\minute} of the experiment with (left) $\mathrm{Ra}=\num{0.5e6}$ and (right) $\mathrm{Ra}=\num{1.5e6}$.
    The red rectangular indicates the re-circulation area, where en enlargement is shown in \autoref{fig:ulm_rbc_top_im}.
    }
    \label{fig:ulm_rbc_bot}
\end{figure}

The exemplary experiments shown here were performed with a \qty{2.5}{\K} and \qty{10}{\K} temperature difference, which corresponded to $\mathrm{Ra}=$ \num{2.5e5} and \num{1e6}, respectively.
A fairly strong magnetic field with the strength of either \qty{100}{\milli\tesla} or \qty{300}{\milli\tesla} was applied, which corresponded to $Q=\num{0.67e5}$ and \num{2e5}, respectively.
Both time-averaged and regular interpolated flow structures are depicted in \autoref{fig:ulm_rbc_bot}.
It is shown that with ULM the first and, partly, the second convection roll were resolved with a maximum penetration depth of \qty{120}{\mm}.
Due to the acoustic directivity pattern of the ultrasound probe, the upper left-hand part and the small lower left-hand part were shadowed.
The upper shadowed region was increased due to the lower displacement of the ultrasound probe.
As expected, the experiment with the lower $\mathrm{Ra}$ incorporated a decreased velocity range.
But compared to earlier works using UDV and UIV \cite{Weik2021b}, a significant higher velocity gradient and vectors closer to the wall can be observed.
As a result, the deformation of the convection roll from an elliptic to a more rectangular shape was shown.
The velocity distribution of the main roll in $\mathrm{Ra}=\num{2.5e5}$ appeared to be more regular with a decreased velocity gradient and circular streamlines almost into the center.
ULM showed experimentally that the zero velocity region was larger with an increased $\mathrm{Ra}$.

\subsection{Re-circulation Areas}
\begin{figure}[!t]
    \centering
    \includegraphics{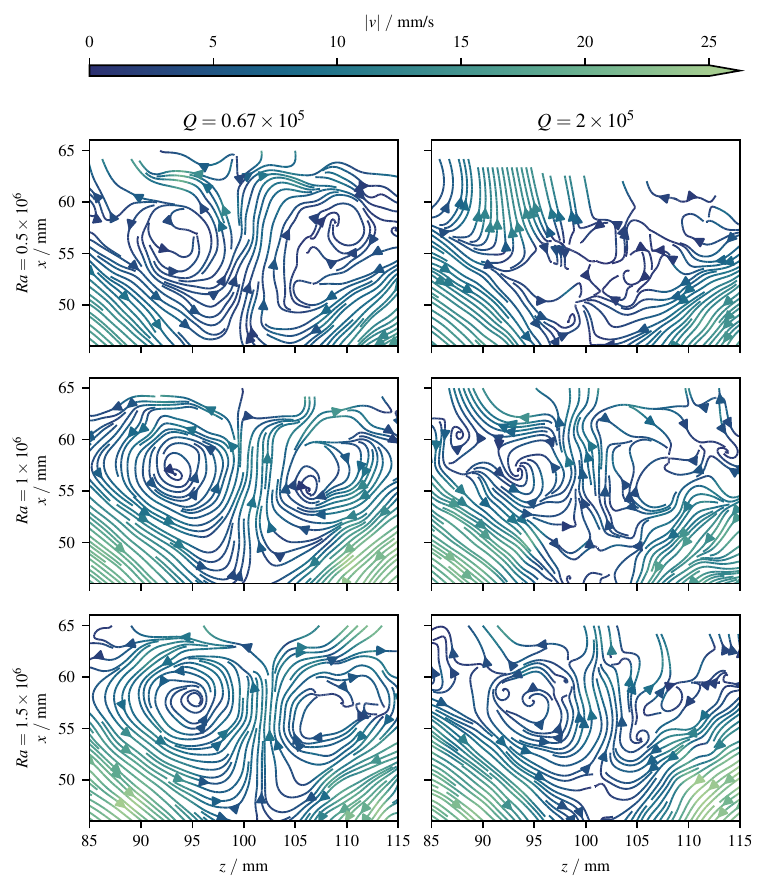}
    \caption{
    Formation of the re-circulation area depending on the environmental parameters $Q$ and $\mathrm{Ra}$.
    }
    \label{fig:ulm_rbc_top_im}
\end{figure}

An enlarged section of the re-circulation area for the experiments in the parameter range is shown in \autoref{fig:ulm_rbc_top_im}.
These results are the first spatially resolved re-circulation areas in this parameter range to be obtained directly from liquid metal experiments.
Here, the sub-millimeter spatial resolution enables to investigate flow gradients and the shape of the re-circulation rolls.
A characterizing value is the ratio of the rotation magnitude between re-circulation area and main vortex.
The values obtained from the experiment in the lower left of \autoref{fig:ulm_rbc_top_im} accounted to $\frac{\qty{4}{\mm\per\s}}{\qty{24}{\mm\per\s}} = \frac{1}{6}$.
This ratio changes in respect to the magnetic field strength, as the re-circulation area with increased $Q=\num{2e5}$ is quite different.
It is less pronounced with a reduced velocity and for lower $\mathrm{Ra}=\num{0.5e6}$ not even visible.
As shown with the uncertainty quantification, a limited spatial resolution leads to an underestimation of the velocity due to a velocity gradient.
Accordingly, an uncertainty of approx. \qty{20}{\percent} in that particular re-circulation area can be considered reasonable.
Furthermore, the lower the velocity, the less information can be obtained from ULM as less particles can be detected in the same amount of time.
In this case, the re-circulation is most likely below the sensitivity threshold of ULM and longer measurement times are required to gather sufficient information.
\par
Nonetheless, the experimental outcome even with an uncertainty of approx. \qty{20}{\percent} are still quite an achievement compared to the state of the art.
Therefore, it is shown that ULM is able to validate predictions made by \textcite{Vogt2018b} in respect to an increased uncertainty as reasonable for experimental data.
This can be accomplished by experiments with a further increased parameter range in the future.

\subsection{Vertical Convection}
\begin{figure}[!ht]
    \centering
    \includegraphics[width=0.3\linewidth]{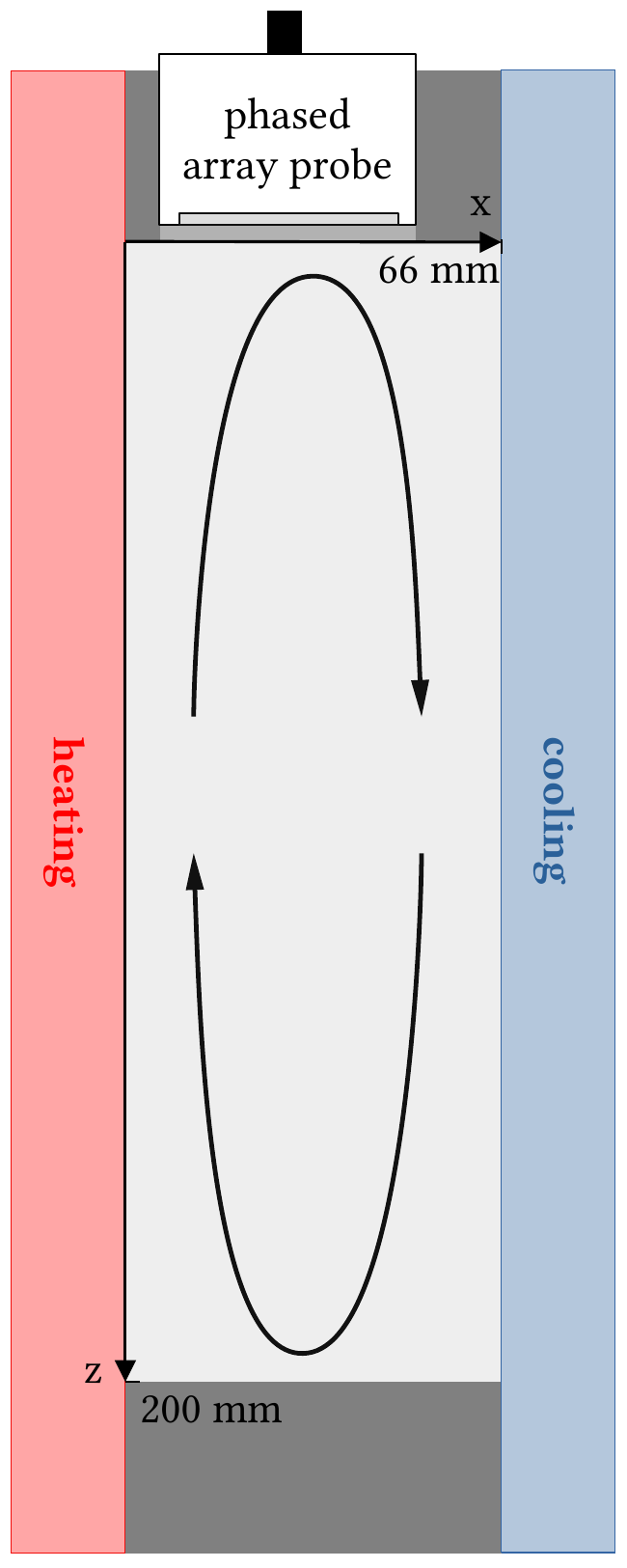}
    \caption[
        Sketch of the vertical convection cell.
    ]{
        Sketch of the flipped setup: Vertical convection cell with a heated wall on the left side and a cooled wall on the right side.
        As with the horizontal convection, the planar flow velocity is measured using the phased array probe attached in between the thermal plates.
    }
    \label{fig:rbc_setup_vertical}
\end{figure}

Unlike the flow confined between horizontal plates that was investigated so far, this section will focus on a vertical convection.
Here, the heating and cooling plates are arranged vertically, which we created by flipping the setup, see \autoref{fig:rbc_setup_vertical}.
This kind of convection is unstable even for the smallest temperature differences.
\par
With this configuration, research is still concerned to identify proper scaling laws from the setup's characteristic Rayleigh number $\mathrm{Ra}$ to the heat transport and boundary layer dimensions.
In \textcite{Zwirner2022} it was shown by simulations and experiments, that the plate size is a much more dominating factor than the aspect ratio in contrast to the horizontal setup.
\par
\begin{figure}[!ht]
    \centering
    \includegraphics[width=\linewidth]{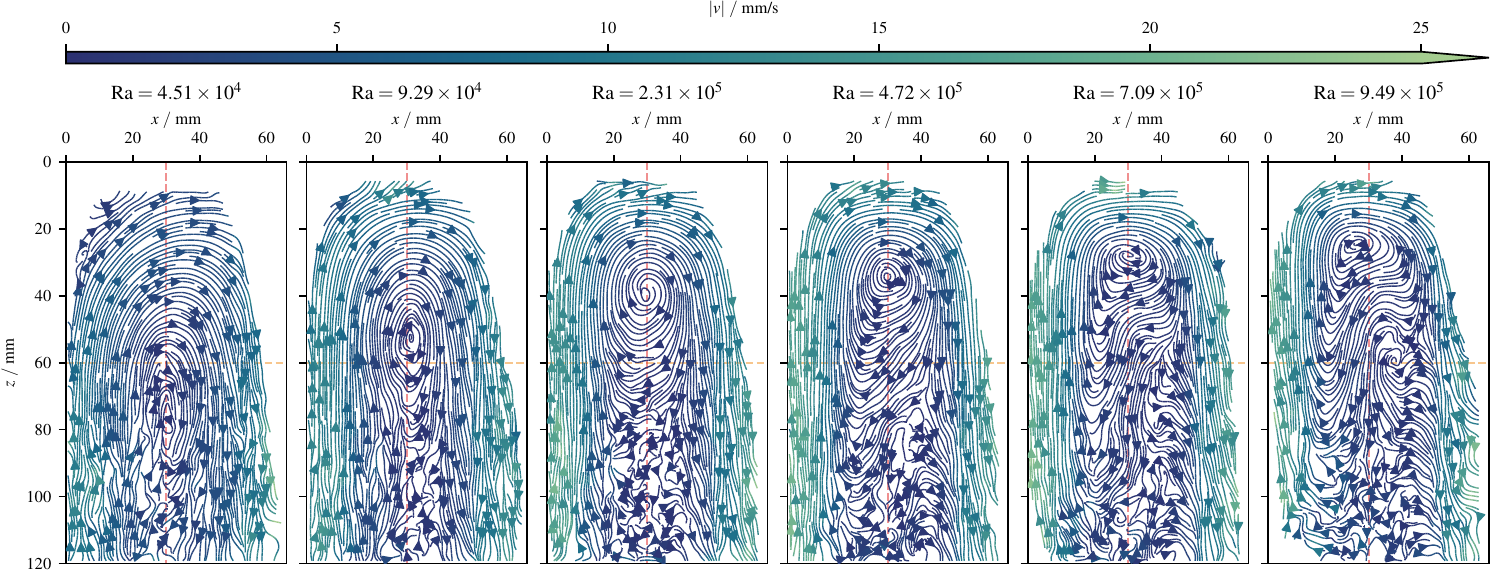}
    \caption{
    ULM flow maps for the vertical convection obtained from an average over \qty{10}{\minute} measurement time.
    The Rayleigh number increases from left to right, which introduces a change of the flow map structure and vortex complexity.
    }
    \label{fig:ulm_rbc_vert}
\end{figure}

Six case study experiments were conducted in the $\mathrm{Ra}$ range of \numrange{4.51d4}{9.49d5}.
The resulting ULM data is visualized with the streamline plots in \autoref{fig:ulm_rbc_vert}.
Here, ULM was able to resolve the scaling of the flow regime according to the increase of $\mathrm{Ra}$ with high detail.
The maximum velocity increases and is further pushed to the walls, i.e. boundary layers.
It starts with a large homogeneous main vortex for low $\mathrm{Ra}$.
For higher $\mathrm{Ra}$ the main vortex is further pushed outwards and smaller vortices in the inner region appear.
As with the horizontal convection, the inner regions are increasingly detached from the main vortex, however, this transition appears with less $\mathrm{Ra}$.
\par
\begin{figure}[!ht]
    \centering
    \begin{subfigure}{0.49\linewidth}
        \includegraphics[]{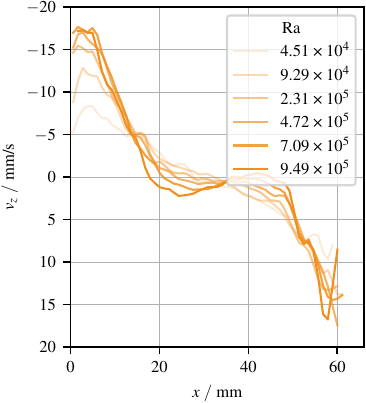}
        \caption{}
        \label{fig:ulm_rbc_vert_cross_a}
    \end{subfigure}
    \hfill
    \begin{subfigure}{0.49\linewidth}
        \includegraphics[]{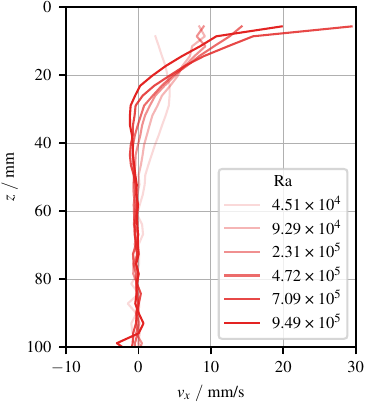}
        \caption{}
        \label{fig:ulm_rbc_vert_cross_b}
    \end{subfigure}
    \caption{
    Velocity profiles at cross-sections indicated in \autoref{fig:ulm_rbc_vert} for increasing $\mathrm{Ra}$.
    (a) velocity in $z$-direction at the line $z=\qty{60}{\mm}$ and (b) velocity in $x$-direction at $x=\qty{30}{\mm}$.
    }
    \label{fig:ulm_rbc_vert_cross}
\end{figure}

The cross-sections of the velocity vectors are able to envision this process with more detail.
In \autoref{fig:ulm_rbc_vert} cross-sections are indicated at $x=\qty{30}{\mm}$ and  $z=\qty{60}{\mm}$, where the corresponding velocity magnitudes $v_x$ and $v_z$ are shown in \autoref{fig:ulm_rbc_vert_cross}.
Interestingly, for $\mathrm{Ra}$ below \num{1d5} the velocity profile in the viscous boundary layer can be detected.
According to \textcite{Zwirner2022} the boundary layer should be in the range of \qtyrange{2.5}{1.5}{\mm}, of which ULM is here shown to detect.
For increased $\mathrm{Ra}$ the boundary layer is further pushed to the wall and the velocity gradient increases.
The increased gradient to the container middle can clearly be observed for these parameters, but to the wall no clear velocity decrease can be observed.
One reason that defines this limitation can be the limited spatial resolution.
The velocity gradient is increased and therefore the uncertainty is too large, c.f. \autoref{eq:unc_gradient}.
\par
Another reason can be that no particles were traced sufficiently close to the wall.
Preliminary experiments demonstrated the ability to detect particles at \qty{100}{\micro\m} distance to the wall \cite{Weik2022}.
However, it can be argued that the values from the static case are not as reproducible here in the dynamic flow setup, where no isolatable echo information was able to be tracked closer than \qty{1}{\mm} to a side wall.
\par
For the lateral velocity in \autoref{fig:ulm_rbc_vert_cross_b} also the same effect is shown: Increased $\mathrm{Ra}$ leads to an increased velocity gradient, velocity maximum and the main vortex closer to the wall.
The main limitation here is given by the PMMA wall: The particle information is shadowed by reverberations.
Even though SVD filtering was applied, no isolatable particle echo was found in this region.

\section{Conclusion}\label{sec:conclusion}
Our results demonstrate that ULM provides high resolution flow maps with \qty{188}{\micro\m} in liquid metal model experiments.
Combining nonlinear adaptive beamforming with typical ULM processing enables super-resolution tracking of melt intrinsic scattering particles.
Trajectories and velocity components can be traced down to a resolution of \num{0.28} wavelengths, an improvement of more than factor 20 compared to UIV approaches \cite{Poelma2017,Weik2021}.
As measurement uncertainty plays an important role in experimental fluid physics, an uncertainty quantification model was presented.
The overall measurement uncertainty of ULM can be estimated by using the count and the standard deviation of the tracking results in combination with a preliminary static calibration for the localization uncertainty.
A typical measurement uncertainty of \qty{4.22}{\percent}.
Still, in the boundary layer region an increased uncertainty of approx. up to \qty{20}{\percent} must be considered as the velocity gradient drastically increases and less trajectories can be detected.
\par
To demonstrate the applicability, a case study of a liquid metal convection experiment was conducted.
This posed an interesting research application for ULM, as it firstly enabled an experimental investigation of the boundary layer velocity.
Furthermore, for extended parameter ranges the present limitations of ULM in this application were revealed.
The spatial resolution for increased velocity gradients and velocity vectors closer than \qty{1.5}{\mm} to the side wall were mainly limiting.
\par
This study will help for the design and target of future works, where we want to combine ULM boundary layer measurements in larger setups as in \textcite{Schindler2022}.
We therefore plan to directly immerse the transducer probe into the melt in the follow up experiments.
This will avoid the information loss due to reverberation.
Furthermore, the sensitivity for particle echoes should further increase as no attenuation and reflection from a window is in the acoustic path.
It is a question to be investigated in the future if this also allows to detect particles closer to the side wall.

\section*{Acknowledgment}
The authors gratefully acknowledge the support of the Deutsche Forschungsgemeinschaft (DFG) under grant 512483557 (BU2241/9-1) upon which this study is based.
\par
We thank S. Singh for his work on the model experiments, S. Su for fruitful discussions to the topic as well as O. Rothkamm and J. Weber for their work on the calibration measurements.

\printbibliography

\end{document}